# Simple route to Nd:YAG transparent ceramics

Yu. A. Barnakov*, I. Veal, Z. Kabato, G. Zhu, M. Bahoura, M. A. Noginov
*Center for Materials Research, Norfolk State University, 700 Park Avenue, Norfolk, 23504 VA*

We report on the fabrication and spectroscopic characterization of transparent $Nd^{3+}$:YAG ceramic, a prospective material for future laser applications.

## INTRODUCTION

Neodymium doped Yttrium Aluminum Garnet (Nd:YAG) has proven to be one of the best solid-state laser materials in the history of quantum electronics. Its indisputable dominance in a broad variety of laser applications is determined by a combination of high emission cross section with long spontaneous emission life-time, high damage threshold, mechanical strengths, high thermal conductivity and consequently low thermal distortion of the laser beam, *etc*. The fact that the Czochralski crystal growth of Nd:YAG is a matured, highly reproducible and relatively easy technological procedure adds significantly to the value of this material. One of the few drawbacks of Nd:YAG is its relatively low growth rate, ~1 mm/hour, making the production of large laser rods and slabs, with the linear size comparable to 10 inches or more, which are needed for high-power and high-energy laser applications, prohibitively expensive.

The later disadvantage stimulated tremendous efforts for alternative synthesis methods. The pioneering research of several groups in Japan in the last decade [1-4] has led to the development of the unique technique allowing one to synthesize highly transparent polycrystalline ceramics of YAG and several other laser materials [5].

Large ceramic laser elements can be produced at relatively low cost, they are free of internal stress or intrinsic birefringence, and allow relatively large doping levels or optimized custom-designed doping profiles. This makes ceramic laser elements particularly important for high-energy laser applications. Thus, 1.46 kW Nd:YAG ceramic laser has been demonstrated [6]. These ceramics are commercially produced by Konoshima Chemical Co. Ltd. in Japan. At the same time, several research groups in the US and Europe, working on the technology of the same material [7-11], were not able to produce laser ceramics of comparable quality.

## CERAMIC FABRICATION

The research program undertaken in the Center for Materials Research at Norfolk State University is aimed at understanding the fundamental principles underlying fabrication of transparent ceramics. We study the relationship between chemical synthesis, processing, microstructure, and optical properties of the ceramics as well as optimize each step of the processes: the synthesis of nanoparticles, the compaction of a green body, and the vacuum sintering. At the present time, we report our recent breakthrough from translucency to transparency in Nd doped YAG ceramics.

Commercially available powders of nominally 2 at.% doped Nd:YAG nanoparticles were used in the experiments. Various additives, including traditional alkylsilanes (TEOS, APTMS) and polymers (PVA, PVB), taken in concentrations 1-5 wt.%, were employed to modify particles' surface and facilitate the compaction. Green body compaction was performed in several different ways with the use of a uniaxial press (22 MPa), cold isostatic press (CIP, up to 330 MPa), and traditional slip casting approach. In the latter method, the powders were re-dispersed in methanol at the presence of additives and then sedimentated naturally in a plastic vial.

*ybarnakov@nsu.edu.

In different particular experiments, the green body density varied in the range 0.2-0.6 of its maximal value (4.56 g/cm$^3$) characteristic of grown single-crystal YAG [12]. The highest value 56% was obtained for a green body formed *via* precipitation followed by air-drying. The vacuum sintering of compacted pellets was performed in several steps characterized by different durations and temperatures. The objectives of different particular stages were (*i*) the removal of impurities, (*ii*) the reduction of pores, and (*iii*) the grain growth. At the latter stage, the ceramic was processed at 1700-1800$^o$C for few hours at the vacuum better than 10$^{-6}$ torr (10$^{-3}$ Pa).

## CHARACTERIZATION OF OPTICAL PROPERTIES

The logo of Norfolk State University can be clearly seen through the sample of fabricated Nd:YAG ceramic (thickness ~0.3 mm), Figure 1. The Scanning Electron Microscope (SEM) image of the same sample shows granules with the average size 3-5 μm, Figure 2.

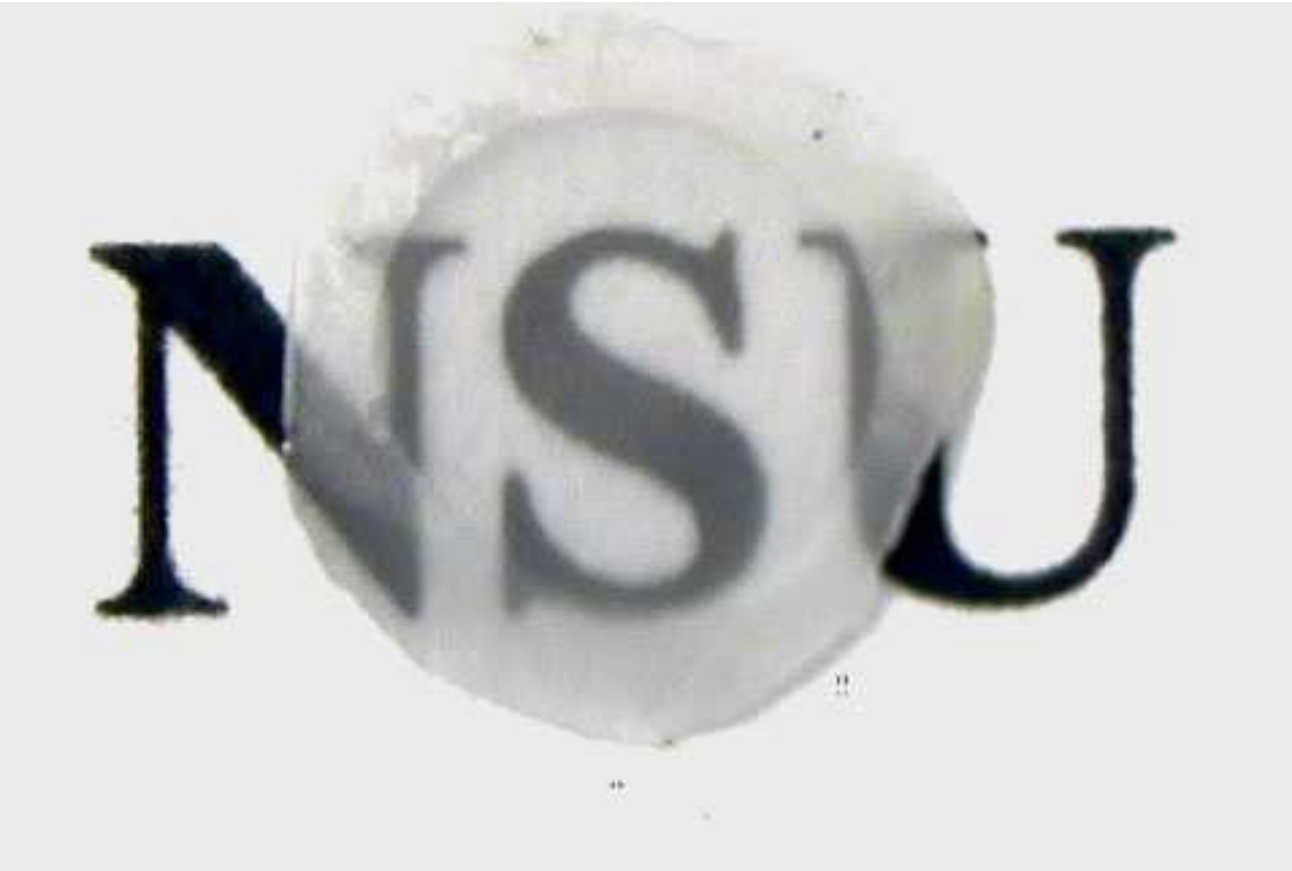

Fig. 1. The photograph of the polished sample of the Nd:YAG ceramics. In the preparation of this particular sample we used 300 nm Nd:YAG powder.

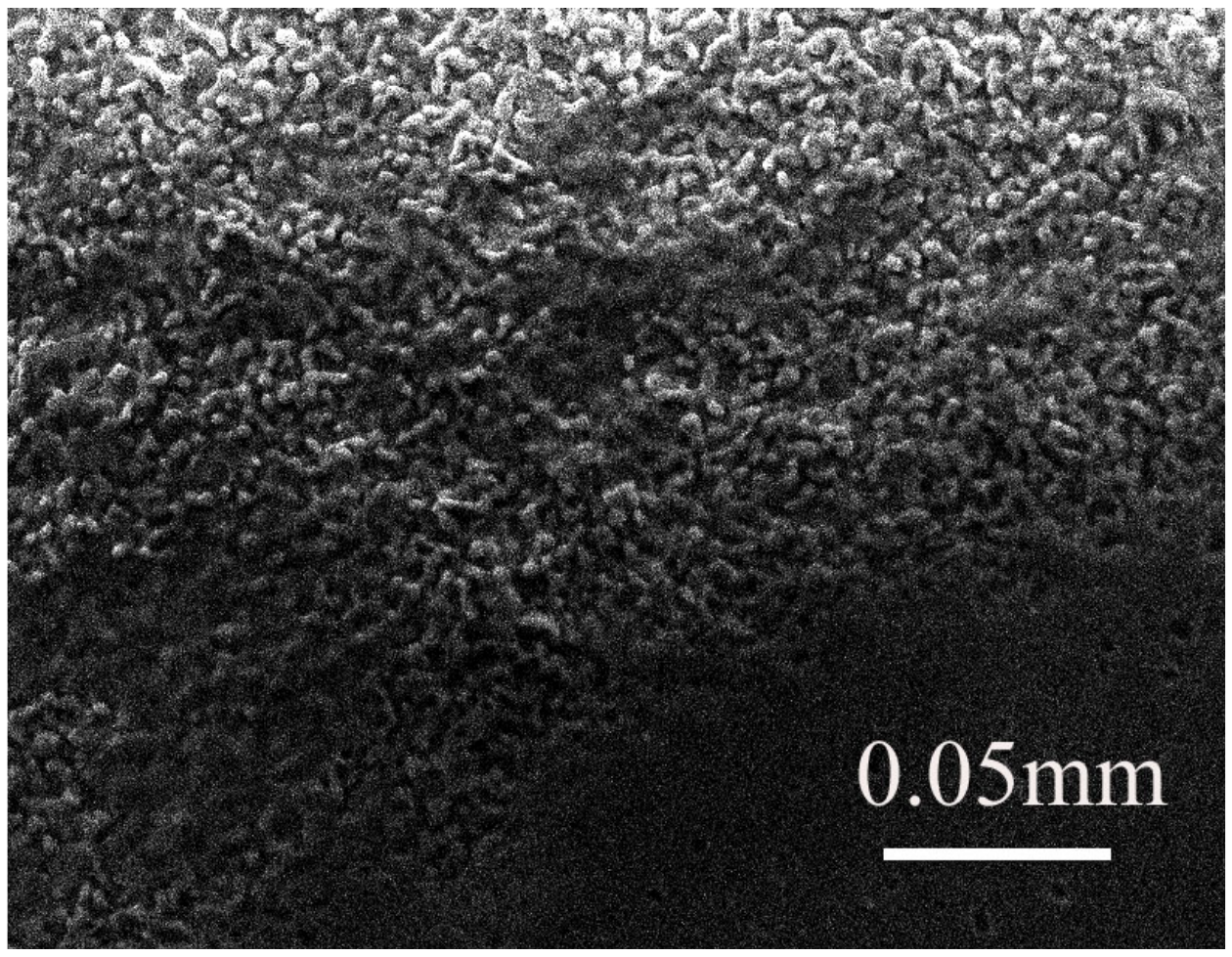

Fig. 2. SEM image of "as-sintered" surface of the Nd:YAG ceramics. (This particular surface was not polished or etched.)

Figure 3 shows the transmission spectrum of the ceramic sample measured in the Perkin Elmer Lambda 900 spectrophotometer using "on-axis" detection without integrating sphere. This measurement allows one to evaluate the scattering length to be equal to $l_s$~ 0.3 mm. The Nd$^{3+}$ absorption peaks are almost identical to those in grown Nd:YAG crystals. However, relatively high intensity absorption line at 588 nm was observed in our ceramic sample. One of the possible explanations of this strong absorption can be the distortion of neodymium sites in the material.

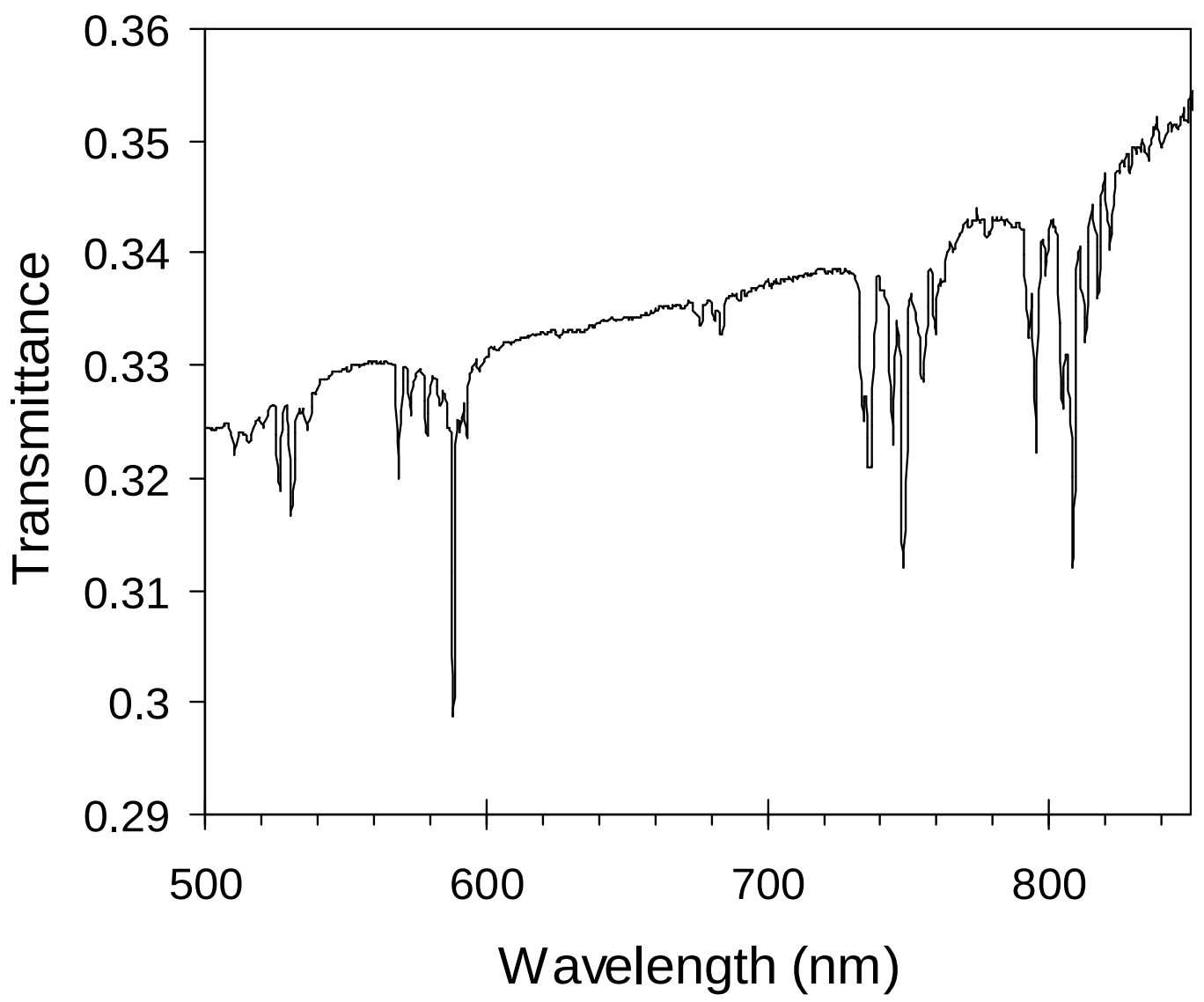

Fig. 3. Transmission spectrum of the Nd:YAG ceramics.

In the emission studies the samples were excited at $\lambda$ = 808 nm with 10 ns pulses of the Optical Parametric Oscillator (Panther pumped by Surelite III from Continuum). The luminescence was recorded using Oriel MS257 monochromator, a photomultiplier tube, and a boxcar integrator. The emission spectrum of the ceramic was very similar to that of the grown crystal, Figure 4. The emission kinetics slightly deviated from single exponential, with the decay time measured at the level 1/e to be equal to 232 μs (the neodymium emission life-time in low doped Nd:YAG crystals is 230 μs [12]).

The reported result is break-through in our research, which gives us hope to obtain good quality transparent ceramics in near future.

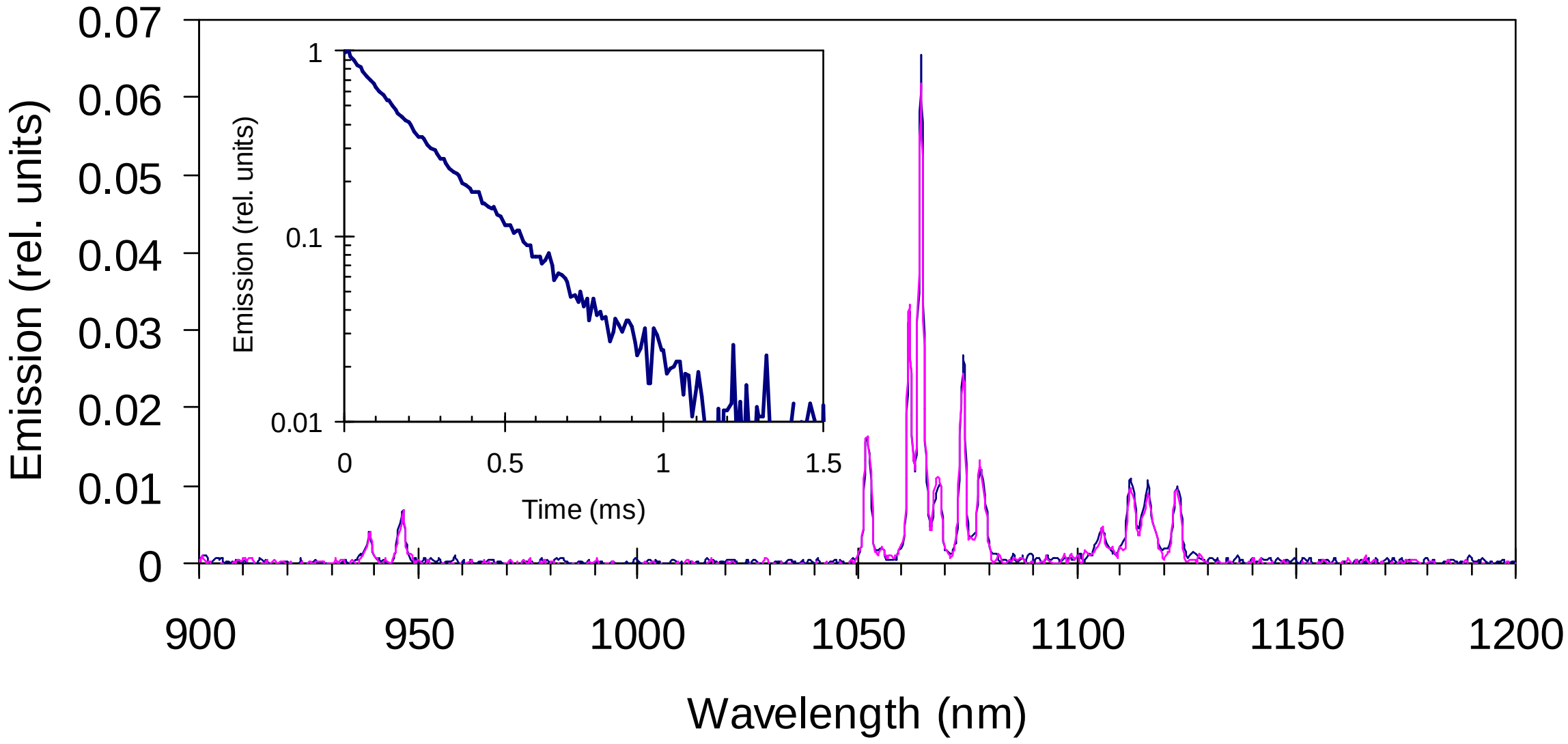

Fig. 4. Overlapping emission spectra of the Nd:YAG ceramic and grown Nd:YAG crystal. Inset – emission kinetics of the Nd:YAG ceramics.

## ACKNOWLEDGMENTS

The authors would like to thank J. Intrater, A. Pradhan and G. Loutts for valuable discussions and the assistance with experiments. The work was partially supported by the NASA grant NCC 3-1035, NSF grant HRD – 0317722, and MDA contract HQ0006-04-C7096.